\documentclass{iopjournal}
\usepackage{amsmath,amssymb,mathtools,bm}
\usepackage{subcaption}
\usepackage{braket}
\usepackage[linesnumbered,ruled,vlined]{algorithm2e} 

\begin{document}

\articletype{Paper} 

\title{Dynamic Depth Quantum Approximate Optimization Algorithm for Solving Constrained Shortest Path Problem}

\author{Rakesh Saini$^{1,*}$\orcid{0000-0002-4963-8465}, Nora Mohamed$^1$\orcid{0009-0001-3095-8709}, Saif Al-Kuwari$^1$\orcid{0000-0002-4402-7710} and Ahmed Farouk $^1$\orcid{0000-0001-8651-9011}}

\affil{$^1$Qatar Center of Quantum Computing, College of Science and Engineering, Hamad Bin Khalifa University, Qatar Foundation, Doha, Qatar.}


\affil{$^*$Corresponding author}

\email{rasa68842@hbku.edu.qa}

\keywords{Quantum Approximate Optimization Algorithm, Constrained Shortest Path Problem, Ising formulation, Quadratic Unconstrained Binary Optimization}

\begin{abstract}
The Quantum Approximate Optimization Algorithm (QAOA) has emerged as a promising approach for solving NP-hard combinatorial optimization problems on noisy intermediate-scale quantum (NISQ) hardware. However, its performance is critically dependent on the selection of the circuit depth—a parameter that must be specified a priori without clear guidance. In this paper, we introduce a variant of QAOA called dynamic depth Quantum Approximate Optimization Algorithm (DDQAOA) that resolves the challenge of pre-selecting a fixed circuit depth. Our method adaptively expands circuit depth, starting from $p=1$ and progressing up to $p=10$, by transferring learned parameters to deeper circuits based on convergence criteria. We tested this approach on 100 instances of the Constrained Shortest Path Problem (CSPP) at 10-qubit and 16-qubit scales. Our DDQAOA achieved superior approximation ratios and success probabilities with fewer CNOT gate evaluations than the standard QAOA for $p=3$, $5$, $10$, and $15$. In particular, while standard QAOA at $p=15$ achieved results close to our approach, it used 217\% and 159.3\% more CNOT gates for 10-qubit and 16-qubit instances, respectively. This demonstrates the performance and practical applicability of DDQAOA to solve combinatorial optimization problems on near-term devices.
\end{abstract}
\section{Introduction}
Combinatorial optimization problems arise in a wide range of applications, from large-scale integrated circuit design and drug discovery to financial portfolio management. These problems are not only of great theoretical interest but also central to numerous applications in real-world decision-making systems, making them fundamental to both scientific research and industrial practice \cite{goto2019combinatorial, tong2021usco, ramos2020metaheuristics}. However, solving these problems on classical computers is difficult due to the exponentially growing solution space and the ubiquity of local optima \cite{wolsey1999integer, arora2009computational}. 

Quantum computing has emerged as a promising approach to achieve exponential computational acceleration by exploiting quantum phenomena \cite{arute2019quantum, daley2022practical, zhong2020quantum}. However, current quantum hardware faces significant limitations in qubit connectivity, error susceptibility, and qubit counts, placing it firmly within the Noisy Intermediate-Scale Quantum (NISQ) era \cite{bharti2022noisy}. Despite these limitations, NISQ devices provide a platform for developing and testing early generations of quantum algorithms that can potentially demonstrate quantum advantages \cite{preskill2018quantum,cerezo2021variational}. 

Among these algorithms is the Quantum Approximate Optimization Algorithm (QAOA) \cite{farhi2014quantum}, which has become one of the most popular approaches for solving combinatorial optimization problems with its hybrid quantum-classical approach making it particularly well-suited for NISQ hardware \cite{farhi2014quantum, blekos2024review, harrigan2021quantum}. However,  as problem complexity increases, achieving enhanced solution quality often requires deeper quantum circuits, making current experimental implementations impractical due to limitations in decoherence times and gate error rates \cite{wurtz2021maxcut, zhou2020quantum}. While QAOA performance is restricted at lower depths, the higher depths needed to reach optimal solutions exceed the capabilities of NISQ devices and experimental techniques \cite{zou2025multiscale, montanez2025toward}. 

A fundamental challenge in QAOA implementation is determining the optimal circuit depth, which is inherently problem-dependent. In particular, fixed-depth QAOA implementations suffer from fundamental underparametrization as problem density increases since the constraint-to-variable ratio induces performance degradation that cannot be overcome without increasing circuit depth. Although fixed circuit depths may exhibit insensitivity to problem size for certain problems, such as MaxCut, the general requirement that optimal depth $p$ must grow with problem size means that fixed-depth approaches either underperform by using insufficient depth or become computationally wasteful by using excessive depth for simpler instances \cite{crooks2018performance, akshay2022circuit, sureshbabu2024parameter,truger2024warm}. 

To address these limitations, several adaptive approaches have been proposed \cite{tyagin2025qaoa,cheng2024quantum,yanakiev2024dynamic}. The most prominent among these is ADAPT-QAOA \cite{zhu2022adaptive}, which provides a systematic method for iteratively constructing problem-tailored ansätze by selecting operators from a predefined pool based on gradient criteria, thereby improving performance while reducing both parameter count and circuit complexity. However, this method suffers from computational overhead due to repeated gradient evaluations at each iteration, where the cost scales with the size of the operator pool. Complementing these advances, parameter transfer methods \cite{montanez2025toward, sureshbabu2024parameter} enable warm-starting QAOA by transferring optimized parameters from smaller or related problem instances to target problems, a process that primarily depends on structural properties such as graph degree and initialization states. Moreover, a layerwise variant of QAOA has been developed \cite{lee2024iterative}, which trains the layer parameters one by one; however, this approach can get stuck in suboptimal solutions due to the restricted search space. Warm-started QAOA \cite{egger2021warm} that initializes the QAOA in a quantum state based on a classical approximate solution, which means the performance depends on the solution provided by classical pre-computation. These approaches address specific aspects of QAOA but do not resolve the challenge of selecting an appropriate circuit depth a priori, which is a critical bottleneck.

\subsection{Contributions}
In this paper, we present the dynamic depth quantum approximate optimization algorithm (DDQAOA); a variant of QAOA that resolves the challenge of pre-selecting a fixed circuit depth, thereby improving both the performance and practical applicability of the algorithm. The proposed approach is not merely an incremental improvement, but a hardware-aware alternative that prioritizes efficiency and robustness in the NISQ era. The key contributions of this work are as follows:
\begin{itemize}
    \item Introducing DDQAOA, a variation of QAOA that automatically determines the necessary circuit depth, starting from $p=1$ and progressively increasing it based on performance convergence.
    \item An efficient parameter transfer strategy that leverages learned parameters from shallower circuits to warm-start and accelerate the optimization process of deeper layers.
    \item A comprehensive demonstration of the method's effectiveness on CSPP, which is an NP-hard problem. A non-trivial benchmark for constrained optimization.
    \item Superior efficiency and approximation ratios compared to standard fixed-depth QAOA at $p=3$, $5$, $10$, and $15$.
\end{itemize}

\subsection{Organization}
The remainder of this paper is organized as follows. Section \ref{Background} provides a review of the foundational QAOA framework and the mathematical formulation of the Constrained Shortest Path Problem. Section \ref{DD-QAOA} presents a detailed description of our DDQAOA. In Section \ref{ExRe}, we describe the experimental design, including the problem instances, baseline methods, and performance metrics used for our evaluation. In section \ref{sec_results}, we present and analyze the results of our simulations, demonstrating the advantages of DDQAOA over the standard QAOA. In Section \ref{conclusion}, we summarize our contributions and outline directions for future research.

\section{Preliminaries}\label{Background}
In this section, we briefly review the QAOA framework, including its mathematical formulation and the role of cost and mixer Hamiltonians in the optimization process. We then present the specific formulation of the constrained shortest path problem and discuss the encoding of the problem as Quadratic Unconstrained Binary Optimization (QUBO) problems suitable for quantum algorithms. 

\subsection{Quantum Approximate Optimization Algorithm}
Many interesting real-world problems can be framed as combinatorial optimization problems \cite{papadimitriou1998combinatorial,korte2008combinatorial}. Consider an optimization problem defined on $N$-bit binary strings $\mathbf{z} = z_{1}z_{2}\ldots z_{N} \in \{+1,-1\}^N$, where the goal is to determine a string that maximize/minimize a given objective function $C(\mathbf{z}) = \Sigma_{j} C_{j} (j)$. The mapping of $C(z)$ to the Ising Hamiltonion $H_C$ will be easy if $C_{j}(z)$ depends on not more than two of the $z_i$, whereas if it depends on three or more $z_i$, $C(z)$ may still be mapped onto the Ising Hamiltonian, potentially at the expense of introducing additional auxiliary variables \cite{chancellor2017circuit}. The Ising Hamiltonian is diagonal in the Pauli-Z basis, and the ground state energy, denoted by $E^{(0)}_{C}$, corresponds (up to an irrelevant constant) to the minimum of $C(z)$. 

In QAOA, the quantum circuit is initialized in the equal superposition of all possible bit-strings $\ket{+}^{\otimes N}$, which is achieved by applying the Hadamard gates $H^{\otimes N}$ to $\ket{0}^{\otimes N}$.
Next, a variational ansatz created using the problem Hamiltonion $H_C$ and a mixer Hamiltonion $H_{M} = \sum^{N}_{j=1} = \sigma^{x}_{j}$ alternately with controlled durations is applied, thus achieving the QAOA state $\psi$: 
\begin{align}
    \ket{\psi_{p}(\gamma, \beta)} = e^{-i\beta_p H_M} e^{-i\gamma_p H_C} \cdots e^{-i\beta_1 H_M} e^{-i\gamma_1 H_C} |+\rangle^{\otimes N},
\end{align}
which is parameterized by $2p$ variational parameters $\gamma_i$ and $\beta_i$ where $i = 1, 2, \ldots, p$ is to be optimized by a classical algorithm. The goal of QAOA is to find such optimal parameters $\gamma^*$ and $\beta^*$, so that the expected value
\begin{equation}
F_p(\vec{\gamma}, \vec{\beta}) = \langle\psi_p(\vec{\gamma}, \vec{\beta})|H_C|\psi_p(\vec{\gamma}, \vec{\beta})\rangle,
\end{equation}
is maximized (or minimized). QAOA optimization typically starts with some initial guess of the parameters and performs simplex or gradient-based optimization to get the optimal parameters for the next optimization step. The expectation value is estimated by repeatedly preparing the state $\ket{\psi_p(\vec{\gamma}^*, \vec{\beta}^*}$ at the optimized parameters. Once the optimizition ends, the approximation ratio at the final optimal parameters is calculated to benchmark the performance of the QAOA:
\begin{equation}
r = \frac{F_p(\vec{\gamma}^*, \vec{\beta}^*)}{C_{\max}}.
\end{equation}
QAOA starts with the lowest energy eigenstate of the mixer Hamiltonian and, according to the adiabatic theorem, the eigenstate should evolve toward the minimum energy eigenstate of the cost Hamiltonian. QAOA is a digitized/Trotterized approximation of adiabatic quantum computation that discretizes the continuous adiabatic evolution into alternating unitary operations. In our work, we look for the ground state of the cost Hamiltonian; hence, we aim to minimize the expected value.


\subsection{Constrained shortest path problem} 
Constrained Shortest Path Problem (CSPP) is a fundamental NP-hard problem, even when considering only a single resource constraint, which involves finding a minimum cost path between two points while satisfying a number of resource constraints \cite{garey2002computers,handler1980dual}. This problem serves as an excellent testbed for QAOA benchmarking. The problem, visualized in figure \ref{fig1}, involves finding a minimum-cost path between a source node $s$ and target node $t$ in a bi-directional graph, subject to resource constraints.

Given a bi-directed graph $G = (V, E)$ with vertex set $V = \{1, 2, \ldots, n\}$ representing possible locations, and
edge set $E = \{(i,j) \mid i,j \in V, i \neq j\}$ representing the transition between any two locations. Each edge ($(i,j) \in E$ ) in the graph is associated with:
\begin{itemize}
\item A non-negative travel cost value $c_{ij} \in \mathbb{R}^+$
\item A non-negative resource consumption value $r_{ij} \in (\mathbb{R}^+)^{M}$ with M being a positive integer denoting the number of different resources (e.g., fuel, time, distance, etc.). In this paper, $M = 1$, which means that the problem we consider here is equivalent to the Weight Constrained Shortest Path Problem \cite{ren2023erca}.
\end{itemize}

Let $v_s$ and $v_t$ denote the initial vertex and target vertex, respectively. Let $P(v_1, v_\ell) := \{v_1, v_2, \ldots, v_\ell\}$ denote a path that consists of a list of vertices with each pair of adjacent vertices $v_k, v_{k+1}$, $k \in \{1, 2, \ldots, \ell - 1\}$ connected by an edge $(v_k, v_{k+1}) \in E$. We refer to $P(v_1, v_\ell)$ simply as $P$ when no confusion arises. For a path $P(v_1, v_\ell)$, let $c(P) := \sum_{k=0}^{\ell-1} c(v_k, v_{k+1})$ represent the path cost, which is the cumulative cost of the edges that are present in the path $P$. Similarly, let $r(P) := \sum_{k=0}^{\ell-1} \vec{r}(v_k, v_{k+1})$ denote the path resource cost, which describes the total amount of resource consumed when moving from $v_1$ to $v_\ell$ along $P$. Finally, let $r_{\text{limit}} \in \mathbb{R}^+$ denote the resource limit.

The objective is to find a path $P$ from the source $v_s$ to the target $v_t$ that minimizes the total cost:
\begin{align}
    \min \sum_{(i,j) \in P} c_{ij} \label{eq:cost_objective}
\end{align}
while satisfying the resource constraint:
\begin{align}
    \sum_{(i,j) \in P} r_{ij} \leq r_{\text{limit}} \label{eq:resource_constraint}
\end{align}

In addition to the straightforward application of CSPP (i.e., shortest path subject to a resource limit), the CSPP and its variants naturally arise as auxiliary problems in column generation schemes for air-cargo route planning \cite{derigs2009new}, flight-planning optimization \cite{graves1993flight}, crew-pairing \cite{lavoie1988new,muter2013solving}, the tail-assignment problem in aircraft scheduling \cite{gronkvist2006accelerating}, day-to-day crew operations \cite{stojkovic1998operational}, and crew-rostering \cite{gamache1999column}. Some variants of the CSPP include non-additive objective functions \cite{reinhardt2011multi}, probability functions defined over the network \cite{bertsekas1991analysis}, forbidden-path constraints \cite{villeneuve2005shortest}, and replenishment arcs that reset resource counters \cite{smith2012solving}.

\subsubsection{Quadratic program of CSPP}
To represent the CSPP as a binary quadratic program, we use decision variables $x_{ij} \in \{0,1\}$ that indicate whether edge $(v_i, v_j)$ is included in the optimal path. If it is 1, then there exists an edge from $i$ to $j$; otherwise, it is 0. The quadratic program of CSPP can be formulated as:
\begin{align}
\text{minimize} \sum_{(i,j) \in E} c_{ij} x_{ij} \label{eq:qp_objective} 
\end{align}
Subject to the following constraints:
\begin{align}
 &\sum_{(s,j) \in E} x_{sj} = 1; \sum_{(i,s) \in E} x_{is} = 0 \quad \forall (i, j) \in V \label{eq:source_constraint}\\
 &\sum_{(i,t) \in E} x_{it} = 1; \sum_{(t,j) \in E} x_{tj} = 0 \quad \forall (i, j) \in V \label{eq:sink_constraint}\\
 &\sum_{(i,j) \in E} x_{ij} = \sum_{(j,i) \in E} x_{ji} \quad \forall i \in V \setminus \{s,t\} \label{eq:flow_conservation}\\
 &\sum_{(i,j) \in E} r_{ij} x_{ij} \leq r_\text{limit} \label{eq:resource_budget}
\end{align}
Eq. \ref{eq:source_constraint} imposes the source constraints, which state that there should be an edge living source and there should be no edges entering the source. Also, Eq. \ref{eq:sink_constraint} applies the target constraints: only one edge can enter the target, and no edge leaves from the target. Moreover, flow conservation constraints are imposed by Eq. \ref{eq:flow_conservation}, stating that for the vertices except $s$ and $t$, the number of edges entering a vertex should be equal to the number of edges leaving the vertex. Finally, Eq. \ref{eq:resource_budget} bounds the optimal path $P$ not to use the resources more than $r_\text{limit}$.
\subsubsection{Mapping to a QUBO}
CSPP cannot be solved directly by QAOA, which are designed to find the ground state of unconstrained combinatorial optimization problems. Therefore, the problem must be transformed into an unconstrained version, which is achieved by reformulating it as a QUBO problem. The QUBO formalism is a cornerstone of both classical and quantum annealing, as well as variational algorithms, and seeks to minimize a function of the form:
\begin{align}
    f(x) = x^T Q x = \sum_{i,j} Q_{ij} x_i x_j
\end{align}
where $x$ is a vector of binary variables and $Q$ is a real-valued square matrix encoding the problem’s structure.

This transformation involves integrating the constraints from Eqs. \ref{eq:source_constraint}, \ref{eq:sink_constraint}, \ref{eq:flow_conservation}, \ref{eq:resource_budget} into an objective function as quadratic penalty terms. This results in a single QUBO objective function, which is a weighted sum of the original cost and the following penalties: 
\begin{align}
H_{\text{CSPP}} &= H_{\text{cost}} + H_{\text{resource}} + H_{\text{flow}} \label{eq:penalty_hamiltonian}\\
H_{\text{cost}} &= \sum_{(i,j) \in E} c_{ij} x_{ij}\\
H_{\text{resource}} &= \rho \Bigg( \sum_{(i,j) \in E} r_{ij} x_{ij} - r_{\text{limit}} \Bigg)^2\\
H_{\text{flow}} &= \lambda \Bigg[
   \Bigg(\sum_{(s,j) \in E} x_{sj} - 1\Bigg)^2
   + \Bigg(\sum_{(i,t) \in E} x_{it} - 1\Bigg)^2 
 + \Bigg(\sum_{(j,s) \in E} x_{js}\Bigg)^2
   +  \Bigg(\sum_{(t,j) \in E} x_{tj}\Bigg)^2 \\ \nonumber
&\quad + \sum_{v \in V \setminus \{s,t\}}
   \Bigg(\sum_{(i,v) \in E} x_{iv} - \sum_{(v,j) \in E} x_{vj}\Bigg)^2
\Bigg]
\end{align}
where $\rho$ and $\{\lambda\}$ are penalty coefficients that have a value of zero when its corresponding constraint is satisfied and a very large positive value (must be set large enough to dominate the cost term) when violated. 

The final step in this mapping is to convert the classical QUBO expression into a quantum mechanical Ising Hamiltonian. This is done by the standard transformation $x_i = (1 - s_i)/2$, where $s_i \in \{-1,+1\}$ are spin variables corresponding to Pauli-$Z$ operators $Z_i$  acting on the $i$-th qubit. This yields the final problem Hamiltonian:

This penalty formulation can be systematically converted to standard QUBO form:
\begin{align}
H_{\text{QUBO}} = \min \quad \mathbf{x}^{\top} Q \mathbf{x} + \mathbf{g}^{\top} \mathbf{x} + c_0 \label{eq:qubo_form}
\end{align}
where the QUBO matrix $Q$, linear vector $\mathbf{g}$, and constant $c_0$ encode the cost function, resource constraints, and flow conservation requirements.

Finally, the QUBO formulation is converted to an Ising Hamiltonian through the standard transformation $x_i = (1 - s_i)/2$ where $s_i \in \{-1,+1\}$ are spin variables corresponding to Pauli-$Z$ operators $Z_i$. This process results in an Ising problem Hamiltonian of the general form:
\begin{align}
H_{\text{Ising}} = E_0 \mathbb{I} + \sum_{i} h_i Z_i + \sum_{i<j} J_{ij} Z_i Z_j \label{eq:ising_hamiltonian}
\end{align}
where the parameters $\{h_i\}$ (local fields), $\{J_{ij}\}$ (couplings), and $E_0$ (constant) are derived from the QUBO coefficients. The ground state of this Hamiltonian corresponds to the optimal solution of the CSPP.
\begin{figure}[!h]
    \centering
    \includegraphics[width=\linewidth]{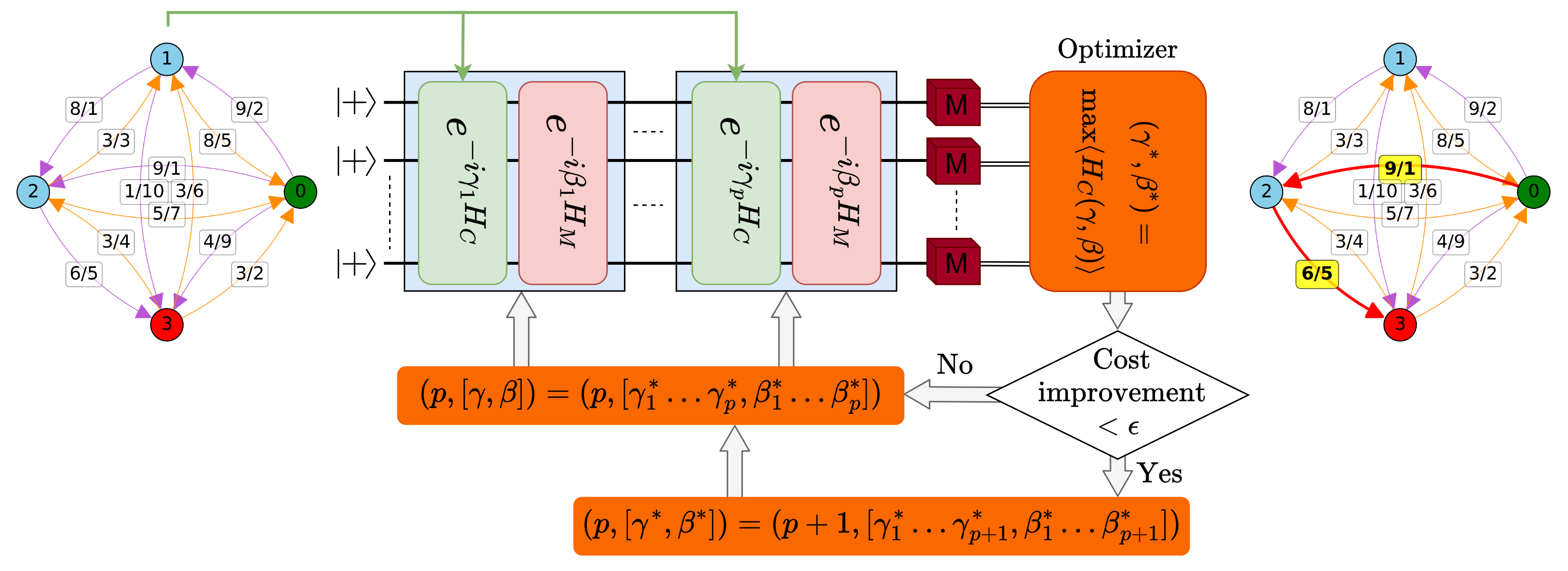}
    \caption{The dynamic depth quantum approximate algorithm (DDQAOA) quantum circuit schematic applied to 4-node CSPP instance. Each box in the circuit contains the cost Hamiltonian operator $e^{i\gamma H_C}$ and mixer operator $e^{i\beta H_M}$, $H_C$ corresponds to the problem. The algorithm iteratively optimize $(\gamma, \beta)$ at depth $p$ using a classical optimizer to maximize $\langle H_C(\gamma,\beta)\rangle$. Here, the DDQAOA starts with $p=1$. When the cost improvement between consecutive iterations falls below threshold $\epsilon$, the depth $p$ is increased by 1 and the optimized $p$ parameters are interpolated to $p+1$ parameters $(\gamma_1^*\ldots\gamma_{p+1}^*, \beta_1^*\ldots\beta_{p+1}^*)$. The solution shown in the right displays the optimal CSPP solution from node $0$ to node $3$, highlighted with a bold red line.}
    \label{fig1}
\end{figure}
\section{Dynamic Depth Quantum Approximate Optimization Algorithm}\label{DD-QAOA} 
In this paper, we propose the Dynamic Depth Quantum Approximate Optimization Algorithm (DDQAOA). Our motivation for developing DDQAOA stems from three observations. First, in fixed-depth QAOA, one must decide how many layers to use. For this, the user tries different numbers of layers until they find an optimal or sub-optimal number of layers. If the number of layers is small, the circuit will be underparameterized and will not converge to an optimal solution. Conversely, if a large number of layers is provided, unnecessary overhead will be introduced \cite{vijendran2024expressive,akshay2022circuit,pan2022automatic}. Second, if the number of layers is fixed throughout the optimization, even with an optimal circuit depth, one has to perform a significant amount of CNOT gate simulation during the optimization. As CNOT gates induce noise, minimizing the number of cost unitaries in the QAOA circuit could be valuable \cite{long2024layering}. Third, the trend of optimal parameters $(\gamma^*, \beta^*)$ should be increasing and decreasing, respectively, according to adiabaticity; however, fixed-depth QAOA yields a random trend in the learned parameters \cite{ruan2025linear}. 

In general, for a given QAOA, the optimal depth can be arbitrary and as $p$ approaches $\infty$, the convergence is guaranteed. Hence, in our DDQAOA, we first initialize the optimization process with a minimum number of QAOA layers, i.e., 1, and then systematically increase the depth when the convergence of the expectation value reaches a stable point. We used a dual convergence detection mechanism. The first mechanism detects a plateau by tracking the expectation value optimization, where the energy improvement falls below a convergence tolerance and no significant improvement occurs for consecutive iterations, indicating the exhaustion of optimization capacity at the current circuit depth. The second mechanism is an enhanced convergence check, which analyzes the variance over recent iterations to distinguish true convergence from oscillatory behavior near local minima. 

Next, when transferring the learned optimized parameter from $p$ to $p+1$, we used an adaptive interpolation method. For $p=1$ to $p=2$:
\begin{align}
\boldsymbol{\gamma}^{(2)} &= [\gamma_1^{(1)}, 1.2 \cdot \gamma_1^{(1)}] \\
\boldsymbol{\beta}^{(2)} &= [\beta_1^{(1)}, 0.8 \cdot \beta_1^{(1)}]
\end{align}

The choice of scaling factors ($1.2$ for $\gamma$ and $0.8$ for $\beta$) is motivated by the adiabatic evolution principle \cite{ruan2025linear}. For $p \geq 2$, the parameters at depth $p$ are mapped to uniformly spaced indices $\{0, 1/(p-1), 2/(p-1), \ldots, 1\}$, and the parameters for $p+1$ are generated by interpolating at indices $\{0, 1/p, 2/p, \ldots, 1\}$. The interpolation method depends on $p$: if $p \geq 4$ use cubic interpolation, otherwise, we use linear interpolation. Algorithm \ref{alg:dynamic_qaoa_corrected} provides a pseudocode for our DDQAOA and figure \ref{fig1} overall quantum circuit schematic for DDQAOA solving an CSPP instance.

\begin{algorithm}[!h]
\caption{The algorithm represent the Pseudo code of DDQAOA.}
\label{alg:dynamic_qaoa_corrected}
\SetAlgoLined
\DontPrintSemicolon

\KwIn{
    $p_0 = 1$;
    $p_{\text{max}}$;
    $H_C, H_M$\;
    $\epsilon > 0$: Energy improvement threshold\;
    $\sigma > 0$: Variance threshold for convergence\;
    $k \in \mathbb{N}$: Patience\;
    $N_{\text{opt\_max}}$;
    $\text{cost\_function}(\boldsymbol{\gamma}, \boldsymbol{\beta})$\;
}

\BlankLine
$p \leftarrow p_0$\;
Initialize $\boldsymbol{\gamma}, \boldsymbol{\beta}$ for $p_0$\;
Initialize optimizer $\mathcal{O}$\;
$E_{\text{hist}} \leftarrow []$, $E_{\text{best}} \leftarrow \infty$, $\boldsymbol{\gamma}_{\text{best}} \leftarrow \boldsymbol{\gamma}$, $\boldsymbol{\beta}_{\text{best}} \leftarrow \boldsymbol{\beta}$\;
$c \leftarrow 0$ \;
\For{$t = 1$ \KwTo $N_{\text{opt\_max}}$}{
    $(\boldsymbol{\gamma}, \boldsymbol{\beta}) \leftarrow \mathcal{O}.\text{step}(\text{cost\_function}, \boldsymbol{\gamma}, \boldsymbol{\beta})$\;
    $E_t \leftarrow \text{cost\_function}(\boldsymbol{\gamma}, \boldsymbol{\beta})$\;
    Append $E_t$ to $E_{\text{hist}}$\;
    
    \uIf{$E_t < E_{\text{best}} - \epsilon$}{
        $E_{\text{best}} \leftarrow E_t$, $\boldsymbol{\gamma}_{\text{best}} \leftarrow \boldsymbol{\gamma}$, $\boldsymbol{\beta}_{\text{best}} \leftarrow \boldsymbol{\beta}$\;
        $c \leftarrow 0$\;
    }\Else{
        $c \leftarrow c + 1$\;
    }
    \If{$c \geq k$ or $\text{Var}(E_{\text{hist}}[-\lceil k//2 \rceil:]) < \sigma$}{
        \uIf{$p = p_{\text{max}}$}{
            \textbf{break} 
        }
        $p_{\text{old}} \leftarrow p$\;
        $p \leftarrow p + 1$\;
        $\boldsymbol{\gamma}_{\text{old}} \leftarrow \boldsymbol{\gamma}_{\text{best}}$, $\boldsymbol{\beta}_{\text{old}} \leftarrow \boldsymbol{\beta}_{\text{best}}$\;
        \uIf{$p_{\text{old}} = 1$}{
            $\gamma \leftarrow [\gamma_{\text{old}}, 1.2 *\gamma_{\text{old}}]$\;
            $\beta \leftarrow [\beta_{\text{old}}, 0.8 *\beta_{\text{old}}]$\;
        }
        \Else{
        $\text{kind} \leftarrow \text{'cubic'}$ if $p_{\text{old}} \geq 4$ else $\text{'linear'}$\;
        $\boldsymbol{\gamma} \leftarrow \text{interpolate}(\boldsymbol{\gamma}_{\text{old}}, p, \text{kind})$\;
        $\boldsymbol{\beta} \leftarrow \text{interpolate}(\boldsymbol{\beta}_{\text{old}}, p, \text{kind})$\;
        }
        Re-initialize optimizer $\mathcal{O}$\;
        $c \leftarrow 0$\;
    }
}
$E_{\text{best}} \leftarrow \text{cost\_function}(\boldsymbol{\gamma}, \boldsymbol{\beta})$\;
$\boldsymbol{\gamma}_{\text{best}} \leftarrow \boldsymbol{\gamma}$, $\boldsymbol{\beta}_{\text{best}} \leftarrow \boldsymbol{\beta}$\;

\Return $(\boldsymbol{\gamma}_{\text{best}}, \boldsymbol{\beta}_{\text{best}})$, $E_{\text{best}}$, $p$;
\end{algorithm}

\section{Evaluation}\label{ExRe}
We carefully and rigorously evaluated the performance of DDQAOA against standard fixed-depth QAOA protocols, ensuring a fair and comprehensive comparison.

\subsection{Experimental setup} 
We created a testbed consisting of 100 randomly generated CSPP instances for 10-qubit and 16-qubit problem sizes. The generation process was designed to produce a diverse set of problems to test the robustness of the algorithms. We then benchmarked DDQAOA against four standard fixed-depth QAOA implementations with circuit depths of $p=3$, $5$, $10$, and $15$. These baselines represent a range of choices that a practitioner might make, from shallow, resource-efficient circuits to deep, expressive ones.

All experiments were conducted using a PennyLane quantum simulator. We used Adam optimizer for the classical optimization loop. We do not consider the run time or the time-to-solution as performance measures since the timing results obtained from the simulator may not be representative of QAOA performed on a real device. 

\subsection{Evaluation metrics} 
We assessed the solution quality using two complementary metrics:
\begin{itemize}
    \item \textit{Success Probability:} The probability of measuring the true ground state (the optimal solution).
    \item \textit{Approximation Ratio (r):} The ratio of the final expectation value to the true minimum eigenvalue of the Hamiltonian, $r=\langle H^*_{\text{CSPP}} \rangle/E_{\min}$, normalized to the range $[0,1]$. An ideal algorithm achieves $r=1$.
\end{itemize}

\begin{table*}[!h]
\centering
\caption{Comparative Performance Summary of DDQAOA versus Fixed-Depth QAOA ($p=3$, $5$, $10$, $15$). The runtime metric is excluded because the simulator timing do not reflect real device performance. We did not perform QAOA optimization on physical hardware. Metrics (success probability) and (approximation ratio) require ground state knowledge and serve primarily for benchmarking; in practical applications, only (expectation value) provides actionable information without prior solution knowledge.}
\label{comparative_performance}
\resizebox{\textwidth}{!}{%
\begin{tabular}{|l|l|l|l|l|l|l|}
\hline
Method & Qubits & Mean Approx. Ratio ($\sigma$) & Median Approx. Ratio & Mean Success Prob. ($\sigma$) & Median Success Prob. \\
\hline \hline
\textbf{DDQAOA} & \textbf{10} & \textbf{0.969 (0.011)} & \textbf{0.973} & \textbf{0.024 (0.025)} & \textbf{0.017} \\ \hline
 $p=15$ & 10 & 0.953 (0.025) & 0.958 & 0.016 (0.032) & 0.004 \\ \hline
 $p=10$ & 10 & 0.937 (0.028) & 0.944 & 0.011 (0.019) & 0.006 \\\hline
 $p=5$ & 10 & 0.912 (0.068) & 0.923 & 0.009 (0.009) & 0.002 \\ \hline
 $p=3$ & 10 & 0.869 (0.119) & 0.892 & 0.010 (0.016) & 0.006 \\
\hline \hline
\textbf{DDQAOA} & \textbf{16} & \textbf{0.990 (0.003)} & \textbf{0.991} & \textbf{0.006 (0.006)} & \textbf{0.004} \\ \hline
 $p=15$ & 16 & 0.985 (0.006) & 0.986 & 0.004 (0.005) & 0.002 \\ \hline
 $p=10$ & 16 & 0.981 (0.007) & 0.982 & 0.004 (0.006) & 0.002 \\ \hline
 $p=5$ & 16 & 0.976 (0.010) & 0.976 & 0.003 (0.006) & 0.001 \\\hline
 $p=3$ & 16 & 0.965 (0.014) & 0.966 & 0.003 (0.006) & 0.001 \\
\hline
\end{tabular}
}
\end{table*}

\section{Results} \label{sec_results}
Table \ref{comparative_performance} provides a summary of our results. In the following, we provide detailed discussions of these results. 

\begin{figure*}
    \centering
    \begin{subfigure}[b]{0.49\linewidth}
        \centering
        \includegraphics[width=\linewidth]{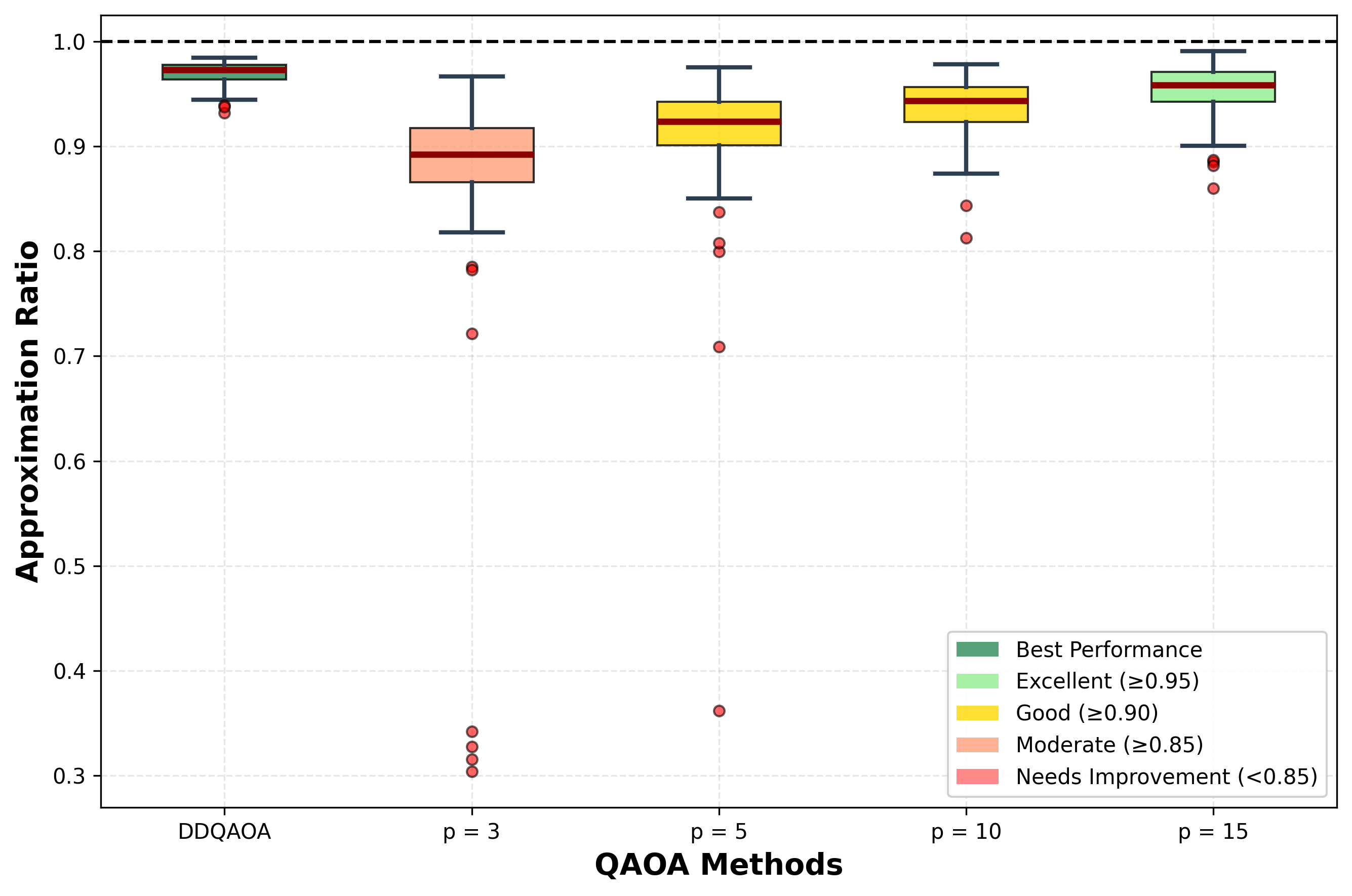}
        \caption{N=10}
        \label{Fig_1a_app_ratio}
    \end{subfigure}
    \hfill
    \begin{subfigure}[b]{0.49\linewidth}
        \centering
        \includegraphics[width=\linewidth]{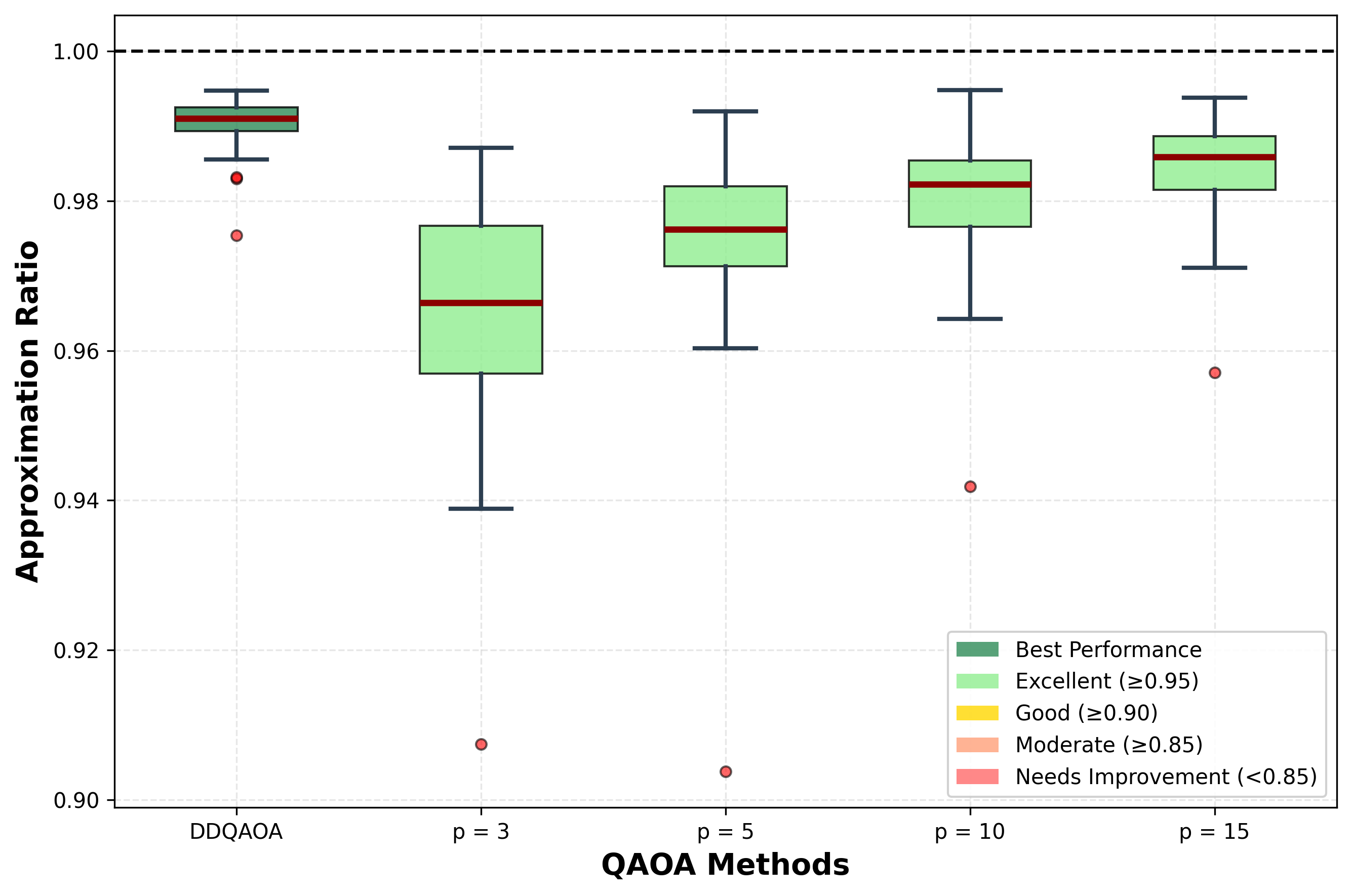}
        \caption{N=16}
        \label{Fig_1b_app_ratio}
    \end{subfigure}
    \caption{The box plot shows the approximation ratio distributions across 100 problem instances achieved by DDQAOA and fixed-depth QAOAs. Black dashed lines indicate the optimal approximation ratio line. DDQAOA substantially exceeds all fixed-depth QAOA results and reached close to the optimal value in all 100 instances.}
    \label{fig:qaoa_analysis}
\end{figure*}

\subsection{Solution quality}
\begin{figure*}[!h]
    \centering
    \begin{subfigure}[b]{0.48\linewidth}
        \centering
        \includegraphics[width=\linewidth]{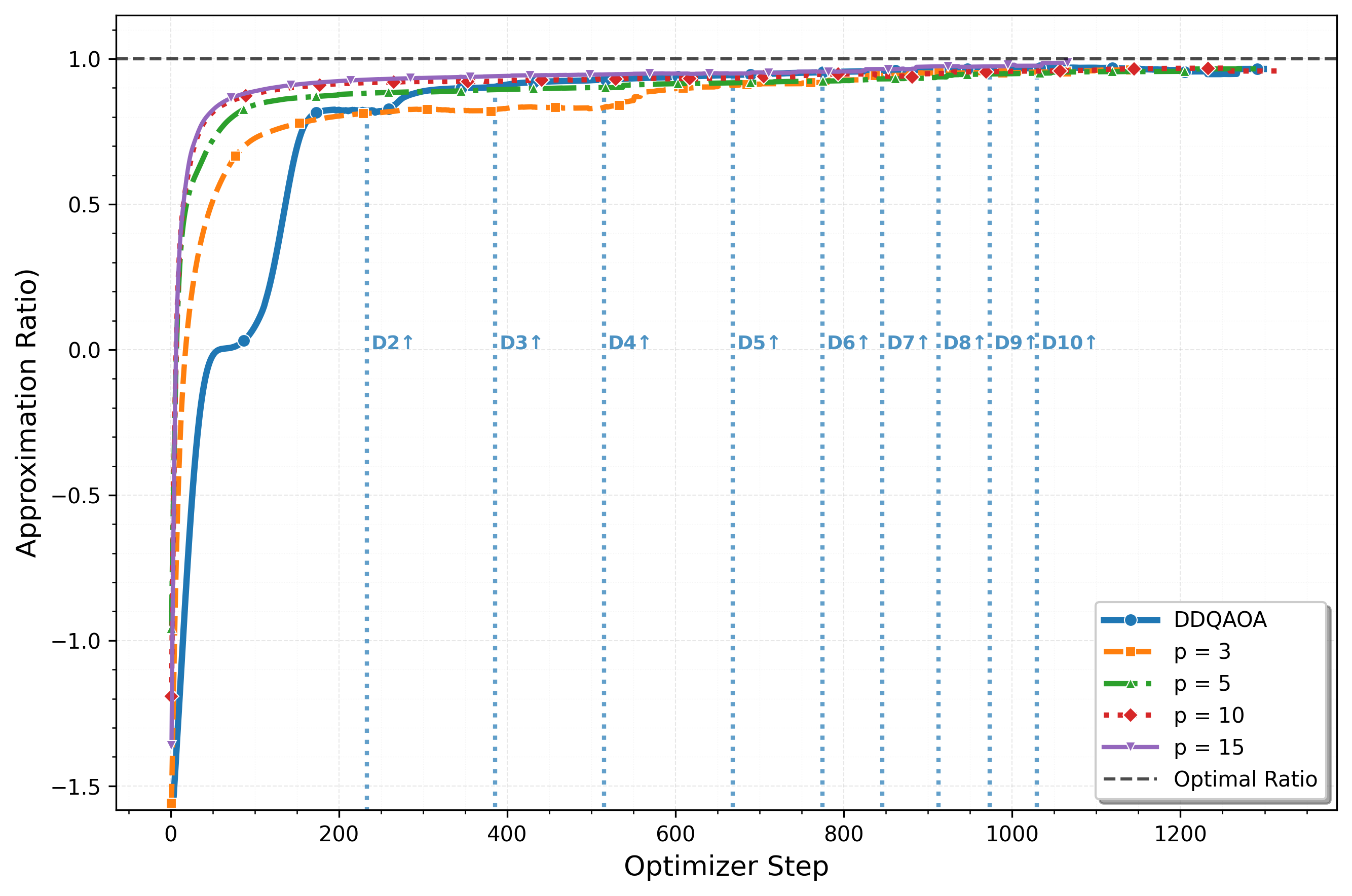}
        \caption{10-qubit approximation ratio}
        \label{fig3a_energy_converge_10}
    \end{subfigure}
    \hfill
    \begin{subfigure}[b]{0.48\linewidth}
        \centering
        \includegraphics[width=\linewidth]{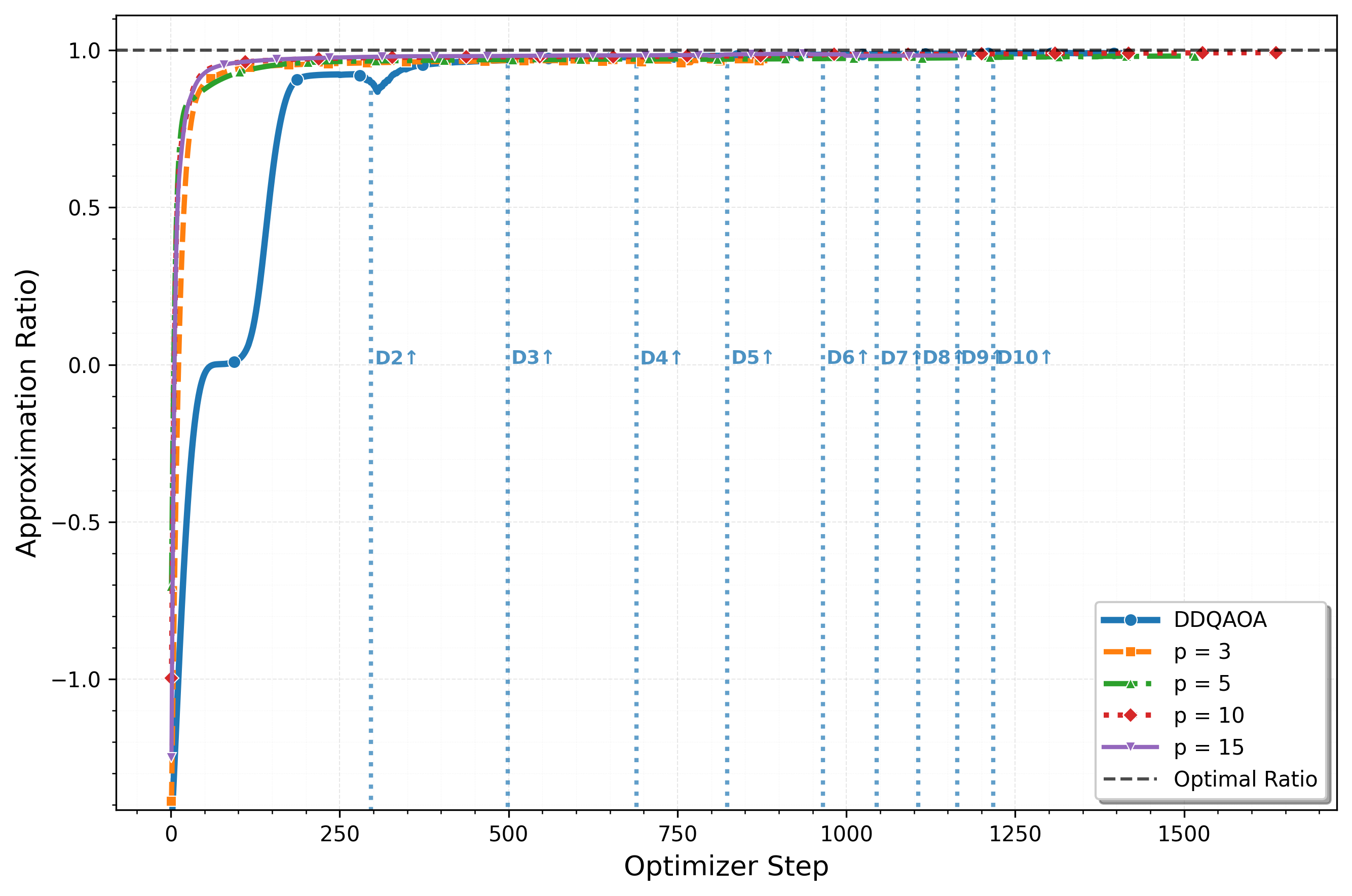}
        \caption{16-qubit approximation ratio}
        \label{fig3b_energy_converge_16}
    \end{subfigure}
    \\
    \vspace{0.3cm}
    \begin{subfigure}[b]{0.48\linewidth}
        \centering
        \includegraphics[width=\linewidth]{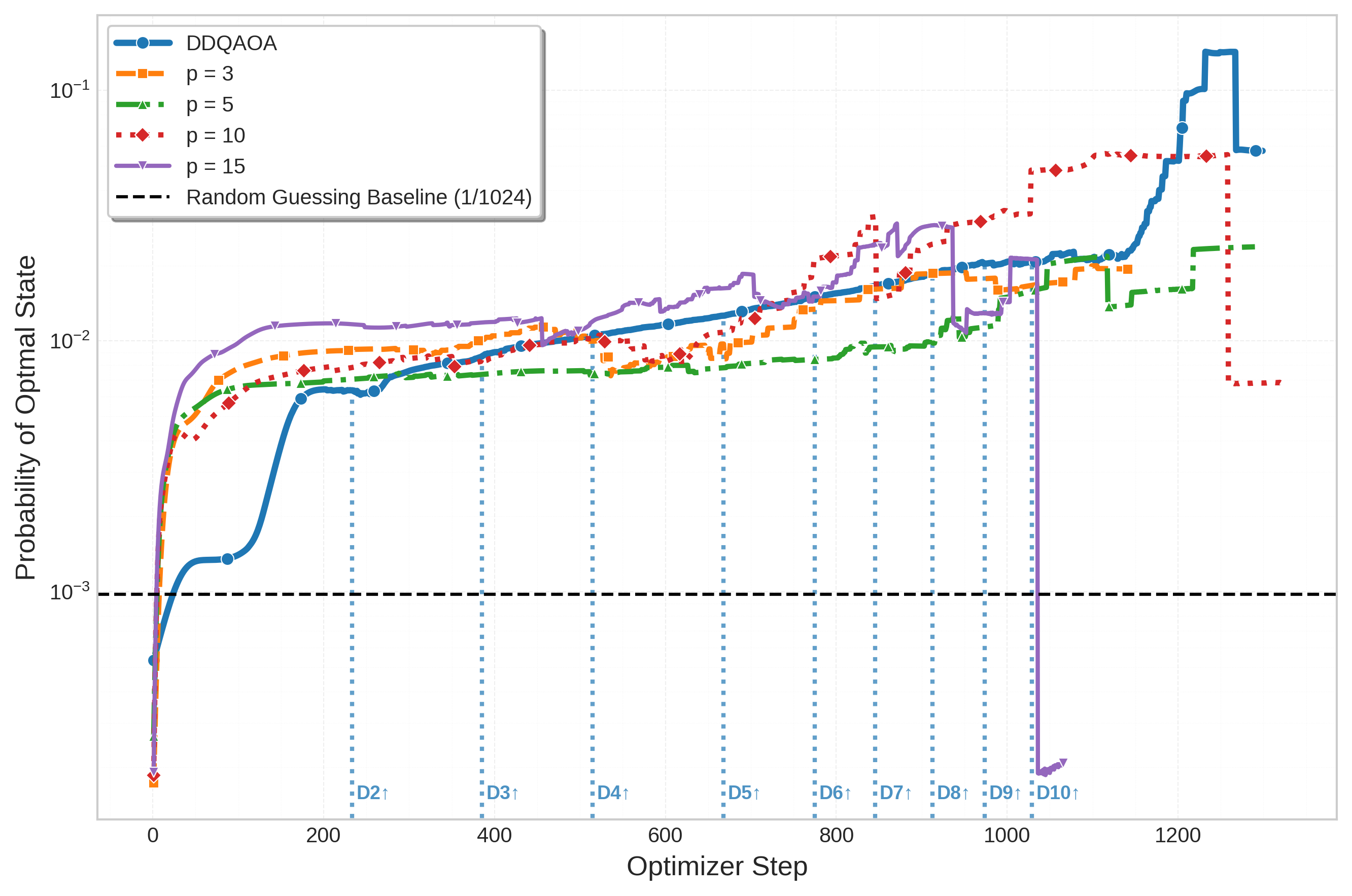}
        \caption{10-qubit success probability}
        \label{fig3c_succ_prob_10}
    \end{subfigure}
    \hfill
    \begin{subfigure}[b]{0.48\linewidth}
        \centering
        \includegraphics[width=\linewidth]{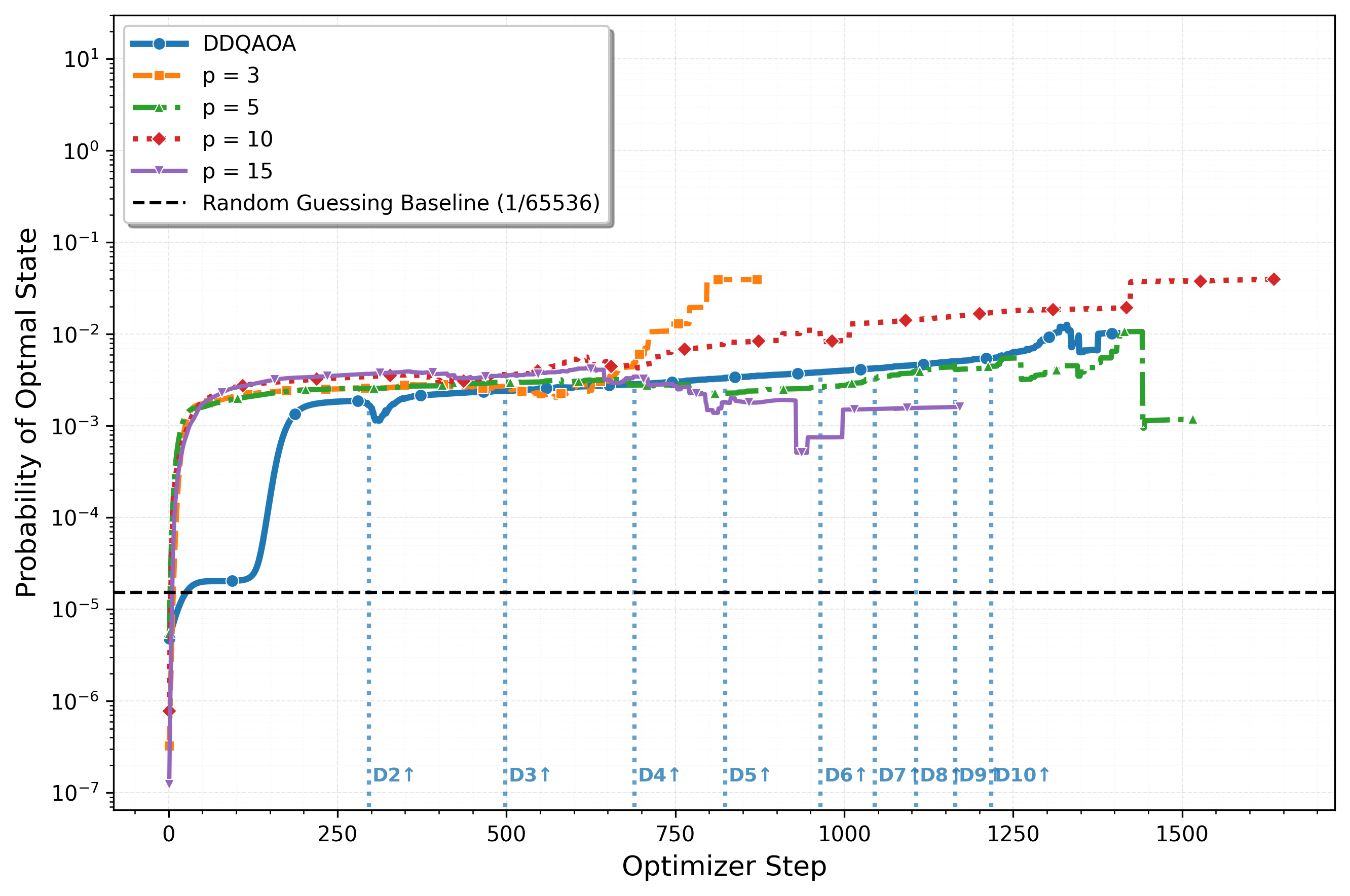}
        \caption{16-qubit success probability}
        \label{fig3d_succ_prob_16}
    \end{subfigure}
    \caption{It shows the convergence trajectories for (a), (b) approximation ratio, and (c), (d) success probability for 10 and 16 qubits. Vertical dashed lines indicate the increase in layer number in DDQAOA throughout the optimization, and the horizontal black dashed line indicates the optimal value for the approximation ratio and the baseline for random guessing success probability.}
    \label{fig:qaoa_convergence}
\end{figure*}

The analysis of solution quality (Approximation Ratio and Success Probability) reveals a clear and consistent advantage for DDQAOA over standard QAOA. figure \ref{fig:qaoa_analysis} shows the distribution of approximation ratios across the 100 test instances for both 10-qubit (figure \ref{Fig_1a_app_ratio}) and 16-qubit (figure \ref{Fig_1b_app_ratio}) problems. For the 10-qubit case, a systematic improvement is observed as the depth of the fixed-layer QAOA increases, with the median approximation ratio rising from 0.89 at $p=3$ to 0.95 at $p=15$. However, DDQAOA surpasses all fixed-depth variants, achieving a higher median ratio of approximately 0.973. More importantly, it exhibits significantly lower standard deviation ($\sigma \approx 0.011$), indicating superior consistency and robustness across different problem instances compared to the much wider spread of shallow circuits, such as $p=3$ ($\sigma \approx 0.119$ with outliers as low as 0.72). Figure \ref{fig3a_energy_converge_10} illustrates convergence throughout the optimization process.

This trend is even more pronounced in the 16-qubit results. While all methods perform well (ratios $\geq$ 0.95), DDQAOA again achieves the highest median approximation ratio ($\approx$0.991) and the smallest interquartile range, surpassing even $p=15$ QAOA. Its standard deviation ($\sigma$=0.003) is evidently lower than that of any fixed-depth method ($\sigma$ from 0.006 to 0.014), confirming its ability to reliably find high-quality solutions. Figure \ref{fig3b_energy_converge_16} shows the convergence.
The results showed that a high-depth QAOA circuit achieves a higher approximation ratio compared to a low-depth QAOA; however, it also requires higher computational resources. DDQAOA evolves the ansatz complexity (increasing $p$) only when the optimization landscape necessitate it, thus the algorithm benefits from growing depth while minimizing computational cost associated with pre-set high depth.

\begin{figure*}[!h]
    \centering
    \begin{subfigure}[b]{0.49\linewidth}
        \centering
        \includegraphics[width=\linewidth]{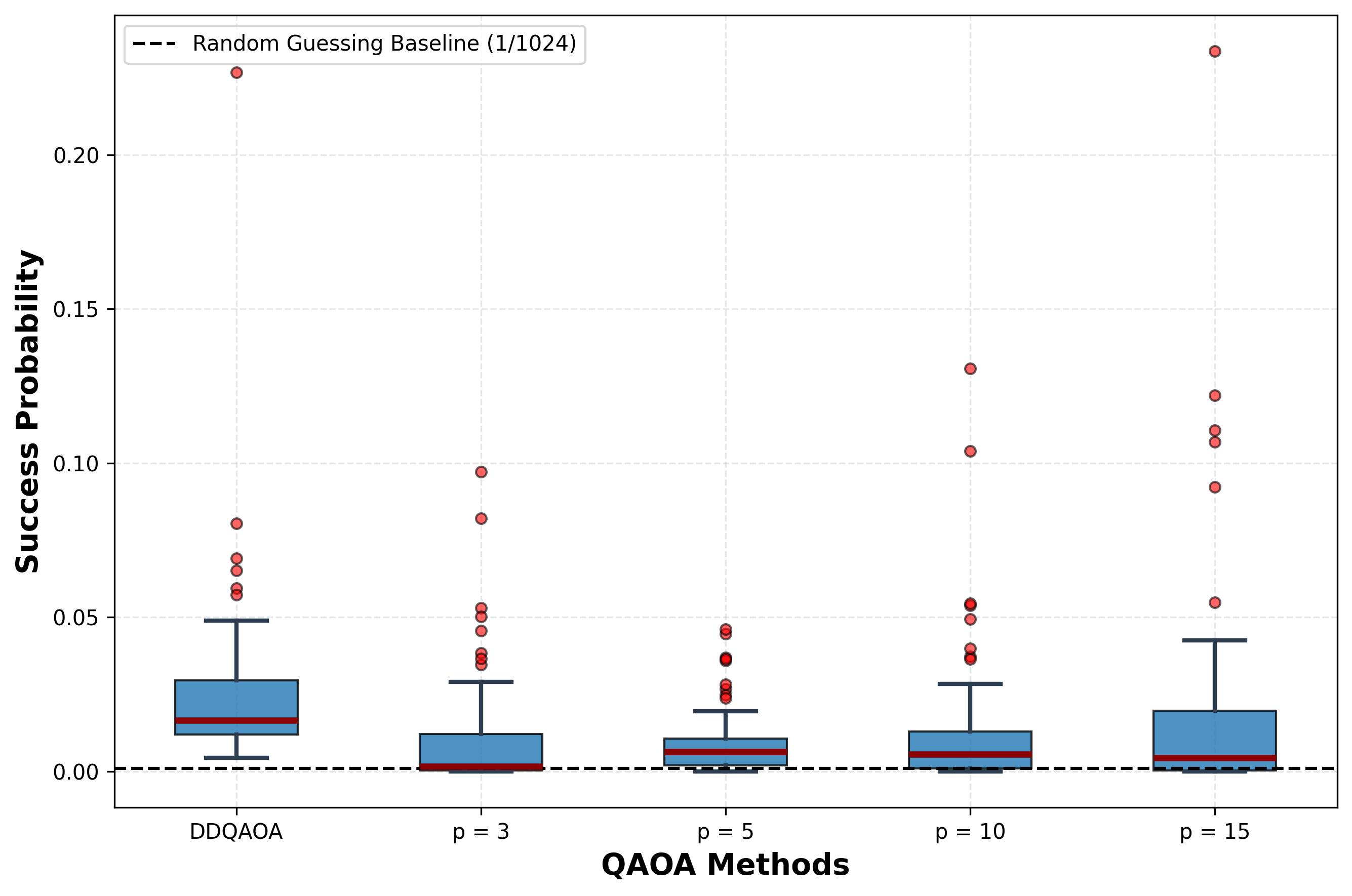}
        \caption{$N=10$ qubits}
        \label{Fig_2a_Succ_prob}
    \end{subfigure}
    \hfill
    \begin{subfigure}[b]{0.49\linewidth}
        \centering
        \includegraphics[width=\linewidth]{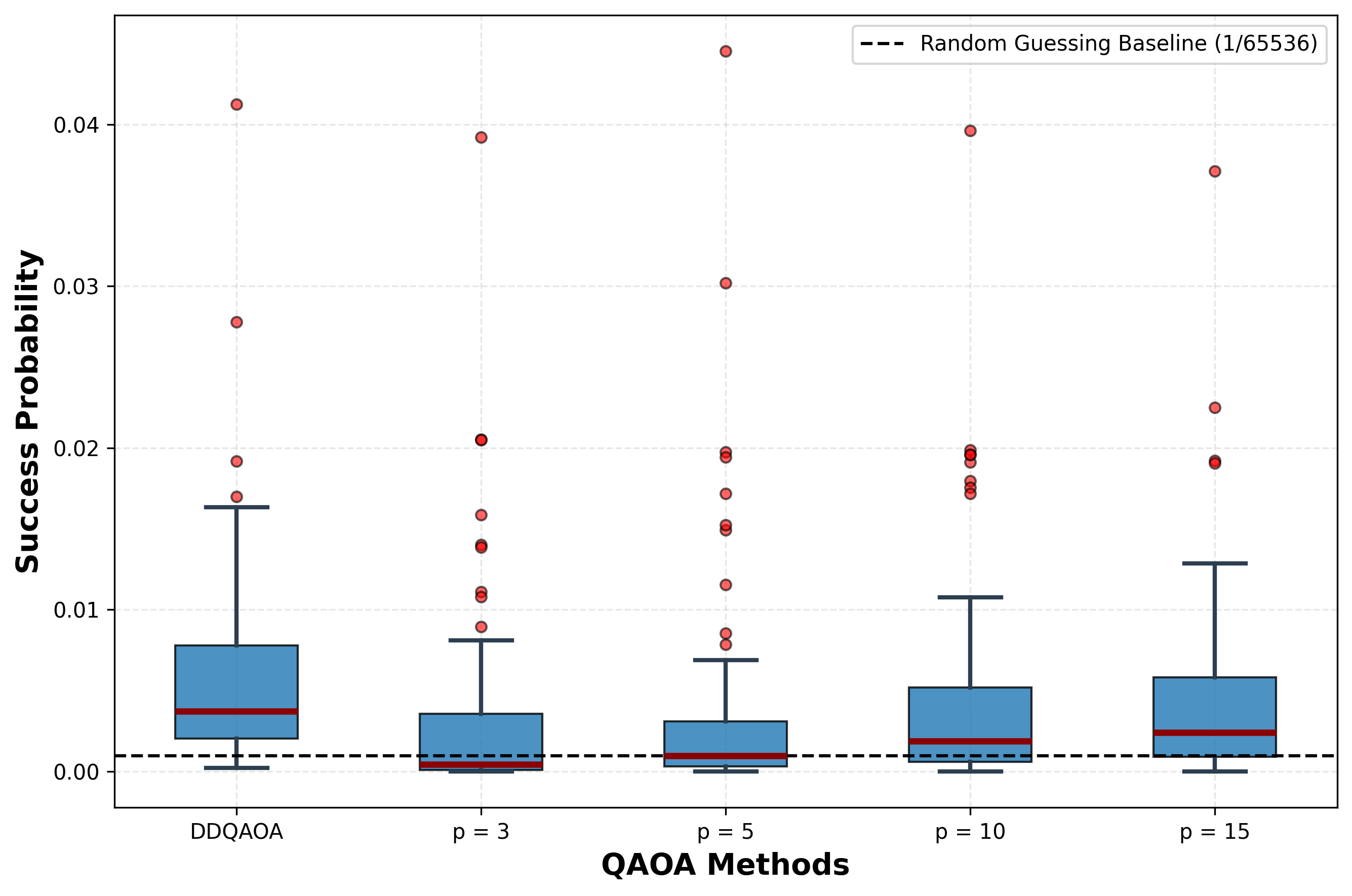}
        \caption{$N=16$ qubits}
        \label{Fig_2b_Succ_prob}
    \end{subfigure}
    \caption{The box plot shows the success probability distributions across 100 problem instances for DDQAOA and fixed-depth QAOAs. Black dashed lines indicate a random guessing baseline. DDQAOA substantially exceeds random performance as well as fixed-depth QAOA results.}
    \label{fig:qaoa_success_prob}
\end{figure*}

The success probability metric, shown in Figure \ref{fig:qaoa_success_prob}, demonstrates a similar performance. For both 10-qubit (figure \ref{Fig_2a_Succ_prob}) and 16-qubit (figure \ref{Fig_2b_Succ_prob}) problems, DDQAOA achieves a substantially higher median success probability (for 10-qubit 0.017 and 0.004 for 16-qubit) than any fixed-depth protocol (for 10-qubit from 0.004 to 0.006 and for 16-qubit from 0.001 to 0.002). This indicates a greater capacity to find the exact ground state, a critical advantage for applications where optimality is required. DDQAOA consistently produces fewer extremely low-value results, demonstrating superior median performance, consistency, and peak success rates. Convergence is shown in Figures \ref{fig3c_succ_prob_10} and \ref{fig3d_succ_prob_16}.
\subsection{Resource cost}
\begin{figure*}
    \centering
    \begin{subfigure}[b]{0.49\linewidth}
        \centering
        \includegraphics[width=\linewidth]{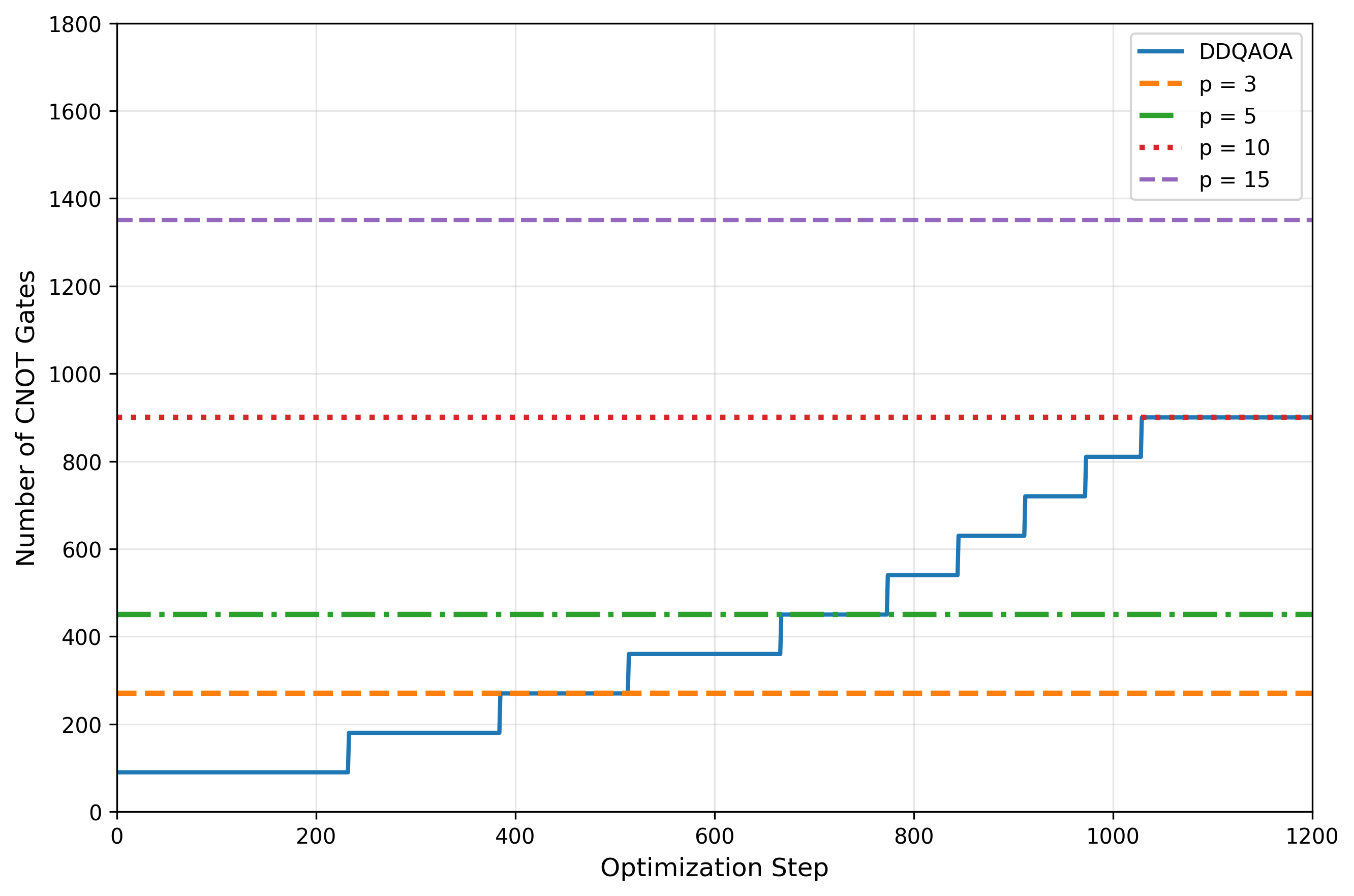}
        \caption{$N=10$ qubits}
        \label{Fig_a_cnot_analysis}
    \end{subfigure}
    \hfill
    \begin{subfigure}[b]{0.49\linewidth}
        \centering
        \includegraphics[width=\linewidth]{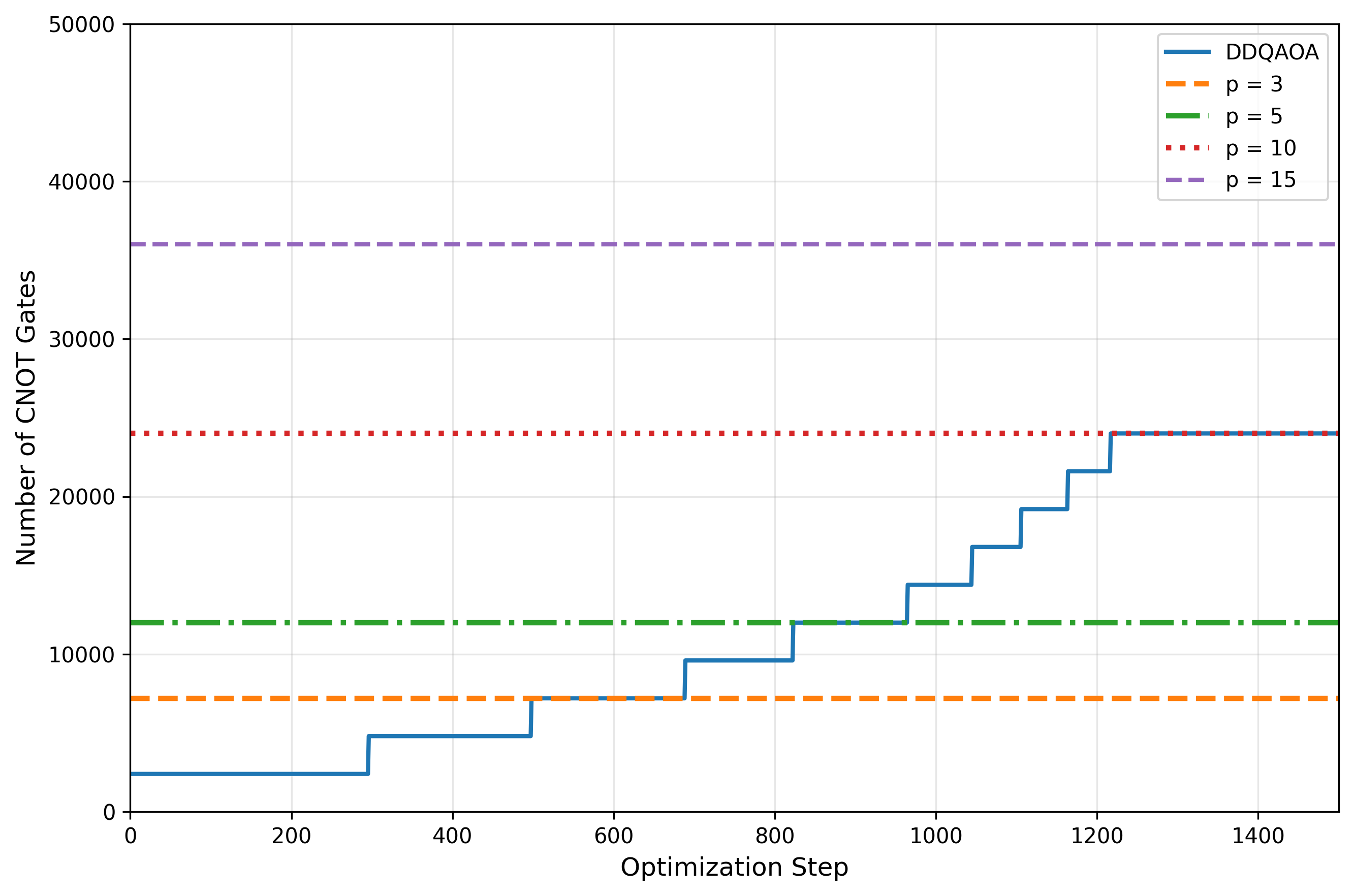}
        \caption{$N=16$ qubits}
        \label{Fig_b_cnot_analysis}
    \end{subfigure}
    \caption{It illustrates the number of CNOT gates in the QAOA circuit at each optimization step. The fixed-depth QAOAs show a constant CNOT utilization throughout the optimization; the DDQAOA shows an increase in the number of CNOT gates, starting from a minimum and reaching equal to the number of CNOT gates utilized by $p=10$ QAOA.}
    \label{fig:cnot_analysis}
\end{figure*}

Figure \ref{fig:cnot_analysis} demonstrates the CNOT gate evolution across optimization steps for both DDQAOA and fixed-depth QAOA implementations. For the 10-qubit instance, DDQAOA increases CNOT gate counts from 90 to 900, and the fixed-depth QAOA uses 270, 450, 900, and 1,350 CNOT gates for $p=3$, $5$, $10$, and $15$, respectively. Similarly, the 16-qubit shows DDQAOA CNOT gates scaling from 2,400 to 24,000, whereas fixed-depth QAOA uses 7,200, 12,000, 24,000, and 36,000 CNOT gates for $p=3$, $5$, $10$, and $15$, respectively. 

DDQAOA begins with minimal circuit resources and incrementally adds QAOA layers throughout the optimization process, thereby increasing the CNOT gate count as needed. This makes DDQAOA more suited for NISQ devices, where shallow circuits minimize error accumulation from gate errors and qubit decoherence. The stepwise expansion pattern shows that the number of CNOT gates scales proportionally to the need, thereby avoiding resource waste by allocating only the necessary circuit depth. In contrast, fixed-depth QAOA variants maintain a constant gate count regardless of solution quality or convergence status. Conversely, DDQAOA prioritizes solution accuracy by expanding QAOA circuit layers on demand, based on convergence behavior. 

From a cumulative resource perspective, the 10-qubit DDQAOA consumes 511020 CNOT gates (gate count multiplied by optimization steps). Compared to this baseline, $p=3$ requires 36.6\% fewer cumulative gates (324,000), $p=5$ uses 5.7\% more gates (540,000), $p=10$ uses 111.3\% more gates (1,080,000), and $p=15$ uses 217\% more gates (1,620,000). For the 16-qubit case, DDQAOA uses 2,082,720 CNOT gates, whereas $p=3$ uses 48.1\% fewer gates (1,080,000), $p=5$ requires 13.6\% fewer gates (1,800,000), $p=10$ uses 72.9\% more gates (3,600,000), and $p=15$ uses 159.3\% more gates (5,400,000).

\subsection{Parameter evolution}
\begin{figure*}[!h]
    \centering
    \begin{subfigure}[b]{\linewidth}
        \centering
        \includegraphics[width=\linewidth]{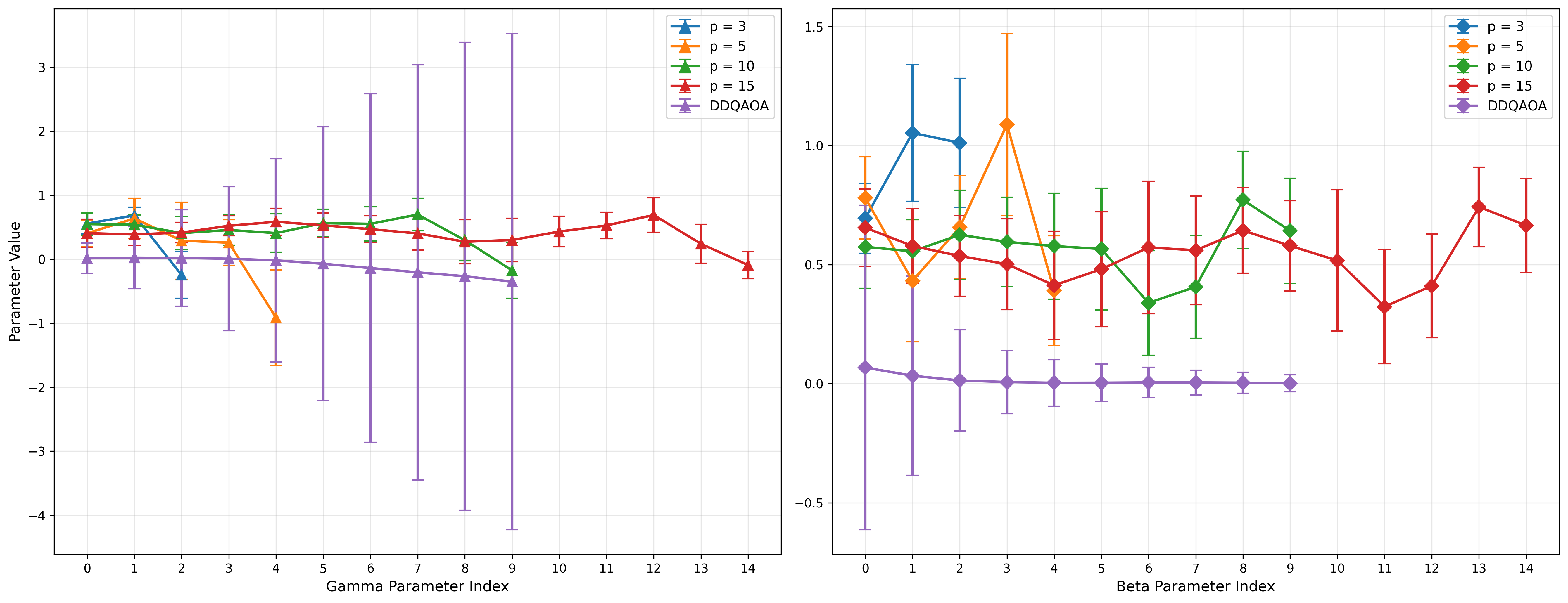}
        \caption{$N=10$ qubits}
        \label{fig4a_final_params_10}
    \end{subfigure}
    \vspace{0.3cm}
    \begin{subfigure}[b]{\linewidth}
        \centering
        \includegraphics[width=\linewidth]{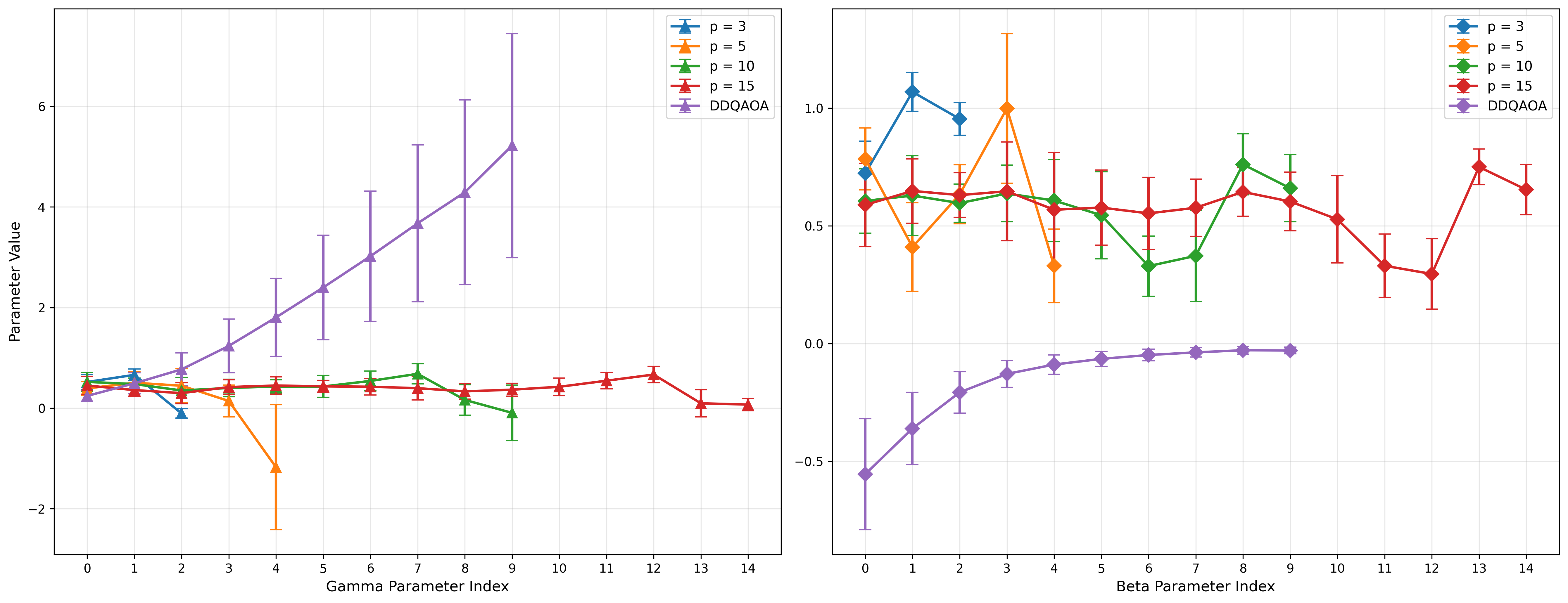}
        \caption{$N=16$ qubits}
        \label{fig4b_final_params_16}
    \end{subfigure}
    \caption{Optimal parameter profiles for DDQAOA and standard QAOA. Left panel shows $\gamma$ parameters (cost Hamiltonian) and right panels for $\beta$ parameters (mixer Hamiltonian). Error bars represent standard deviation across problem instances. The trend of DDQAOA parameter is similar to that of adiabatic theory.}
    \label{fig:qaoa_params}
\end{figure*}

Figures \ref{fig4a_final_params_10} and \ref{fig4b_final_params_16} show the final optimal $\gamma^*$ of cost and $\beta^*$ of mixer Hamiltonian for both 10-qubit and 16-qubit instances, respectively. For fixed-depth QAOA ($p=3$, $5$, $10$, $15$), the optimized parameters exhibit a seemingly random pattern across the parameter indices. In contrast, DDQAOA exhibits monotonically increasing trends across parameter indices that align closely with predictions from adiabatic theory. The $\gamma^*$ parameters show a monotonically increasing pattern across parameter indices, with values consistently growing from near-zero at initial layers to higher magnitudes at later layers. This parameter value behaviour aligns with the adiabatic evolution, where the cost Hamiltonian coefficient should increase toward the last layer ($p-1$) to prioritize exploitation of the cost Hamiltonian. Conversely, $\beta^*$ shows a convergence to zero trend, decreasing from higher initial values to approximately zero at the final layer. This pattern also aligns with the adiabatic conditions that promote exploration in the initial layers, while minimizing mixing in later layers to preserve the evolved state in low-energy configurations near the cost Hamiltonian eigenspace \cite{wurtz2022counterdiabatic}. These structured parameter patterns are consistently observed across both 10-qubit and 16-qubit instances, demonstrating the robustness and scalability of DDQAOA \cite{ruan2025linear}. The adherence of DDQAOA parameters to adiabatic principles suggests that the algorithm effectively approximates the optimal adiabatic passage, achieving parameter configurations that would typically require extensive optimization in standard QAOA formulations.

\section{Conclusion}\label{conclusion}
The quantum approximate optimization algorithm (QAOA) has been extensively utilized to solve various problems, and numerous variants of QAOA have been proposed to enhance its performance. In this paper, we introduce dynamic depth QAOA (DDQAOA), an optimization QAOA variant that iteratively increases the QAOA depth based on the cost convergence. We tested the effectiveness of DDQAOA on the Constrained Shortest Path Problem (CSPP), an NP-hard problem, and tested several instances of it. To examine the performance of DDQAOA, we compared the results against the fixed-depth QAOA with $p=3$, $5$, $10$, and $15$. 

The experimental results provide evidence for the superiority of DDQAOA over fixed-depth QAOA for solving CSPP. Across all evaluated metrics the DDQAOA consistently outperforms or matches the best-performing fixed-layer variants. The method's key advantage lies in its ability to autonomously assign QAOA-layer when needed, avoiding both the underparametrization of shallow QAOA circuits and the wasteful overhead of unnecessarily deep ones. The significantly lower standard deviation in approximation ratios across 100 diverse problem instances demonstrates that DDQAOA is a reliable algorithm, crucial for practical applications.

A particularly compelling finding is the monotonic increase in $\gamma^*$ and the decrease in $\beta^*$  learned by DDQAOA, which can be interpreted as a discretized version of adiabatic evolution. This provides powerful physical intuition for the algorithm's effectiveness: rather than searching randomly through parameter space, DDQAOA discovers a smooth, efficient "annealing schedule" that guides the quantum state from the initial uniform superposition towards the ground state encoding the optimal solution. 

Furthermore, DDQAOA enhances optimization efficiency, converging to high-quality solutions with fewer QAOA layers, resulting in a lower total CNOT gate count. This reduction in quantum resources makes the algorithm more NISQ-aware. By eliminating QAOA depth tuning while improving performance, DDQAOA represents a significant step towards making QAOA a more robust and practical tool for real-world combinatorial optimization.

While these results are promising, the experiments were conducted on classical simulators for systems up to 16 qubits. Future work will focus on solving larger problem instances and validating the performance on real quantum hardware to assess noise resilience and connectivity constraints. Additional algorithmic refinements could also be explored, such as alternative convergence criteria, more sophisticated parameter transfer techniques, and generalization of DDQAOA to other combinatorial optimization problems.


%
%




\data{The data and code that support the findings of this study will be available on reasonable request.}


\bibliographystyle{iopart-num} 
\bibliography{ref}

\end{document}